\documentclass{aa}
\usepackage{natbib}
\usepackage{graphicx}
\newcounter{sub}
\newcounter{subeqn}[sub]
\newcommand\be{\begin{equation}}
\newcommand\ee{\end{equation}}
\newcommand\lp{\left(}
\newcommand\rp{\right)}

\newcommand\st{\stepcounter{sub}}

\newcommand\bea{\begin{eqnarray}}
\newcommand\eea{\end{eqnarray}}
\newcommand\bean{\begin{eqnarray*}}
\newcommand\eean{\end{eqnarray*}}

\newcommand\OOmega{\mbox{\boldmath $\Omega$}}
\newcommand\B{{\bf B}}

\newcommand\no{\nonumber}
\begin{document}
\thesaurus{07     
              (
)}
\title{
Large frequency drifts during Type I X-ray bursts
}
   \author{Vahid Rezania }
   \offprints{V. Rezania\\ email: vrezania@phys.ualberta.ca\\
                  $^*$ Present address 
}
    \institute{Institute for Advanced Studies in Basic Sciences,
               Gava Zang, Zanjan 45195, Iran\\
    Department of Physics, University of Alberta,
              Edmonton AB, Canada T6G 2J1$^*$   
                    }
   \date{Received / accepted }
\maketitle
\begin{abstract}
We study the spin-down of a neutron star atmosphere 
during the Type I X-ray burst in low mass X-ray binaries.  Using polar cap 
acceleration 
models, we show that the resulting stellar ``wind'' torque on the
burning shell due to the flowing charged particles (electrons,
protons and ions) from the star's polar caps 
may change the shell's angular momentum during the burst.
We conclude that the net change in the angular momentum of the star's
atmosphere can account for rather large frequency drifts
observed during Type I X-ray burst.
\end{abstract}
\section{Introduction}\label{int}
It is believed that thermonuclear flashes in pure helium or mixed
hydrogen/helium layers on the surface of weakly-magnetic
($B\leq 10^{10}$ G) accreting (at rate $10^{-11} M_\odot$ yr$^{-1}$
$< \dot{M} < 10^{-8} M_\odot$ yr$^{-1}$) neutron stars cause the observed 
Type X-ray bursts 
in low mass X-ray binaries (LMXBs) \citep{HV75,MC77,LPT95,WT76,B98}.
The accumulated hydrogen and helium on the surface of the neutron star
by accretion ignites and burns periodically.   General observed features of
Type I X-ray bursts, such as burst fluorescences ($\sim 10^{39}$ erg), burst
recurrence times ($\sim$ few hours), and burst durations ($\sim 10-100$ s);
are successfully explained by spherically symmetric thermonuclear runaway
models \citep{LPT95,B98}.  Recent observations from ten LMXBs with the
Rossi X-ray Timing Explorer have revealed nearly-coherent oscillations
during Type I X-ray bursts with frequencies in the range
$\nu_0\sim 270-620$ Hz
\citep{Kli00,Str01}.
Such highly coherent oscillations are interpreted
as rotational modulation of surface brightness asymmetries: 
burning starts locally at some points, creates a brightness
asymmetry, and spreads over the entire stellar surface as the star rotates.
This can be understood by noting that the timescales for accretion between
bursts
is much longer than the burst duration, so the existence of identical
conditions over the whole surface of the star for the burning instability
to start simultaneously is very unlikely.  Therefore,
these oscillations provide a direct measurement of the neutron
star spin frequency \citep{Str96}.  Further, observing such oscillations 
supports the idea that the neutron stars in LMXBs are the
progenitors of the millisecond radio pulsars, see \cite{Bha95} for
a review.

The discovery of such high coherence, large modulation amplitudes,
and stable frequency oscillations while providing the convincing
evidence that the burst oscillation
frequency is related to the neutron star spin frequency \citep{Str01},
however, has brought many new puzzles.
An initial puzzle seen in the observations was that the oscillation
frequency increases by $\Delta \nu\sim$ a few Hz
during the burst.  \cite{Str97} firstly proposed that this frequency shift is
due to the conservation of angular momentum of the decoupled burning shell
from the neutron star in which the shell undergoes spin changes
as it expands and contracts during
the Type I X-ray bursts.  Motivated by this proposal,
\cite{SM99} and \citet{Mil99,Mil00} modeled the observed frequency evolution
and found that the oscillations are coherent, as expected if rotation is 
the cause.  But the first theoretical investigations on understanding
the problem was done by 
\cite{CB00} who studied the rotational evolution of
the neutron star atmosphere during a thermonuclear burst by considering 
one-dimensional vertical hydrostatic expansion.  Since the nuclear burning
time scale ($>$ fraction of a second) is much bigger than the vertical
sound crossing time scale through the burning layer ($\sim$ microseconds),
the outer layers of the neutron star atmosphere are
always in hydrostatic balance.  So by assuming conservation of the
angular momentum of the shell,
a hot burning layer expands hydrostatically 
by $\Delta R\sim 20$ m \citep{AJ82,HF84,B98},
and lags behind the neutron star by
$\Delta\nu\sim \nu_s(2\Delta R/R)\sim 1 {\rm Hz} (\nu_s/300{\rm Hz})
(\Delta R/20{\rm m})(10{\rm km}/R)$ where $\nu_s$ is neutron star spin
frequency and $R$ the radius.  As the layer cools down and
contracts, the rising oscillations due to a temperature inhomogeneity, 
drifts upward as we seen by $\Delta \nu$ of few Hertz.  
By assuming that the burning shell rotates rigidly, they found rough
agreement with the observed values with fractional frequency shifts
$\Delta\nu/\nu_0\leq 0.8$\%.

However, recent observations suggested that purely
radial hydrostatic expansion
and angular momentum conservation alone cannot account for explaining
rather large frequency drifts ($\Delta\nu/\nu_0\sim 1.3$\%) observed  
in some bursts \citep{Gal01,Wij01}.  As pointed out by \cite{Gal01},
the required expansion by these frequency drifts is 4-5 times larger than
the one predicted by \cite{CB00}, see Table 1.  
In order to achieve the rather large frequency shift, \cite{Cum01} improved
the calculations done by \cite{CB00} by including the
general relativistic corrections for either slowly or rapidly rotating
star.   In agreement with \cite{Abr01}, they found that the general
relativity
has small effect on the angular momentum conservation law ($\sim 5-10\%$).
Comparing with the data, for a rigidly rotating atmosphere,
they obtained that the expected spin-down is a factor of two or three
less than the actual observed values. 
In another attempt, 
\cite{Spi01} considered a two-dimensional hydrodynamics model
that the burning
spots can spread over the neutron star surface.  
Due to the combination of the radial expansion of the burning shell
and rotation of the star,
they proposed that horizontal hydrodynamics flows may arise in the 
neutron star burning ocean during the Type I X-ray burst.
By taking into account the action of the Coriolis force due to the
rapid rotation of the star, they showed that the horizontal flows
may explain
many features of observed bursts such as the very short rise time of X-ray
bursts and the lack of burst oscillations in several bursts.   Further,
they argued that during cooling the hot ashes left by burning front, there
is temperature gradient between equator and pole, increasing toward the
pole, which drives a zonal thermal wind directed {\it backward} to the
neutron star rotation.
\footnote{
Actually they showed that due to the hydrostatic balance the material
will likely
ignite sooner at the equator rather than the pole.  As a result, one
would expect the burning front cools down at the equator firstly.}
They found that {\it only if} the frequency drift due to the
radial expansion is combined with the geostrophic drift caused
by backward zonal flows, one may expect to observe the rather
large frequency drifts of burst oscillations in tails of some bursts.

Previous models are based on the angular momentum conservation of
the expanding/contracting neutron star atmosphere during the Type I X-ray
burst.   Although these models are mostly successful in explaining the
many features of Type I bursts, the rather large frequency drifts
observed in some bursts are still unresolved \citep{SB03}.

Because of no coherent pulsations seen in persistent emission
from the majority of neutron stars in LMXBs,  
it is believed that they are weakly magnetized ($B\leq 10^9$ G).
So, in previous studies the effect of magnetic field of
the neutron star was ignored in the rotational evolution of
burning atmosphere during the X-ray bursts.
In this paper we study the change in the angular momentum of
the burning shell during the Type I X-ray burst.   This study is 
based on polar cap particle acceleration models in the pulsar
polar cap regions.  Due to the star's rotation and magnetic field
an electric field must be induced in magnetosphere, i.e. 
${\bf E}=-(\OOmega_s\times{\bf r})\times{\bf B}/c$, that has non-zero
component along the magnetic field lines ($E_{||}={\bf E}\cdot {\bf B}$)
\citep[see for recent review][]{Mes98}.
$\OOmega_s$ is the angular velocity of the star, and $r$ is a
distance outside of the star.
The parallel electric field accelerates charged particles along the
open magnetic field lines above the polar caps
up to ultrarelativistic energies toward the star's light cylinder
($r_{lc}\Omega=c$).
Here, we argue that, for typical magnetic field $B\sim 10^8$ G,
the net charged particles flowing from star
surface to infinity in the pulsar polar caps, would exert a
stellar wind torque on the burning shell that may 
cause the angular momentum of the shell to change during the burst.
We show that the net change in the angular
momentum of the shell during the burst
can account for rather large frequency drifts
of burst oscillations.

\section{The model}\label{model}

The theory of particle acceleration in pulsar magnetosphere has been
studied for three decades, after pioneering
work by \cite{GJ69} on pulsar electrodynamics.  The charged 
particles accelerated
by the parallel electric field (relative to the magnetic field
lines) escape from the surface along the open field
lines which
extended beyond the star's light cylinder (of radius $r_{lc}\sim
5\times 10^9/\nu_s$ cm, where $\nu_s$ in Hertz), and form a
relativistic stellar wind.
Several kind of acceleration mechanisms has been developed \citep{Mes98},
however, the one proposed by \cite{Aro79,Aro81} may apply likely 
to X-ray burst pulsars that we will consider here.
Arons assumed that charged particles (electrons
and ions) flow freely from the neutron star surface due to the thermal
activities.
Freely escaping particles from the neutron star surface depends on the
binding
energy (referred as the cohesive energy for ions and as the work function
for electrons) and the surface temperature. For typical magnetic field
of a neutron star in LMXBs ($\sim10^{8}$ G), the threshold temperature
for thermionic emission of electrons is
$T_e\sim 10^4$ K and for ions is $T_i\sim 10^3$ K \citep{LSM00} that are
$10^4-10^5$ lower that surface temperature of the neutron star in LMXBs.
Further, due to the thermonuclear activities on the surface of the star
during the X-ray burst, the particles' kinetic energy
would increase significantly.
Therefore, freely flowing particles is very likely from the
surface of such a pulsar.

Outward flowing charged particles above the pulsar polar caps
along the open magnetic field
lines causes the Goldreich-Julian (corotation) charge density
$\rho_{\rm GJ}\sim \OOmega_s\cdot\B/2\pi c$ cannot be kept balanced
\citep{HM98}.
As a result, a strong electric field develops along  
the magnetic field, $E_{||}$, above the
magnetic poles due to the departure of total charge density $\rho_e$ 
from the Goldreich-Julian charge
density, starts from zero at the surface and grows with distance above
the surface.     
Even though $\rho_e=\rho_{\rm GJ}$, and therefore $E_{||}=0$
at the surface, the curvature of the field lines causes the area of the
open-field region to increase, so that $\rho_e$ increases faster than
$\rho_{\rm GJ}$, and then a charge deficit grows with distance
\citep{HM98}.
Calculation of the parallel electric field is so complicated and
depends on
several mechanisms that work to enhance or screen the field.
In a series of papers \citet{HM98,HM01,HM02}; and
\cite{HMZ02} have extensively studied the influence
of pair production, inertial frame-dragging effect, curvature
radiation, and inverse Compton scattering on the $E_{||}$.
They showed that for the unsaturated region with altitude in range
$0<z\ll(\Omega_s R/c)^{1/2}\simeq 0.1$
(for $\Omega_s=600\pi(\nu_s/300~{\rm Hz})$ rad/s and $R\sim 10^6$ cm)
above the stellar surface,
the parallel electric field grows linearly
with height as $E_{||}(\theta, \phi, z)= E_0(\theta,\phi) z$,
while for the saturated region with $z> (\Omega_s R/c)^{1/2}$,
the parallel electric field that is nearly
constant respect to the altitude, drops by three order of magnitudes.
Here $\theta$ and $\phi$ are spherical polar and azimuthal angles,
and $z$ is the altitude in units of stellar radius.
The amplitude $E_0(\theta,\phi)$ depends on
magnetic field strength $B$, spin frequency $\nu_s$, and orientation of the
magnetic field symmetric axis relative to the star spin axis $\chi$.
In a simple form, one can estimate $E_{||}$ when the charged particles flow
freely from the neutron star surface \citep[see][]{Aro81,UM95,HM01} as
\st
\bea\label{E||}
E_{||}\simeq 8.2 \times 10^{5}~{\rm V/m}~~ 
\lp\frac{\nu_s}{300~{\rm Hz}}\rp^{5/2}
\lp\frac{R}{10^6~{\rm cm}}\rp^{5/2}\times\no\\
\hspace{2cm}\lp\frac{B}{10^8~{\rm G}}\rp
 \lp\frac{z}{10^{-4}}\rp.
\eea
Note we calculated $E_{||}$ here for
a typical neutron star in LMXBs with nearly perpendicular field 
orientation relative to the spin axis.
The relative fraction of electric force to the gravity force 
above the burning front during the burst, ie. $z\sim 10^{-3}$,
for electrons and protons are $\sim 8\times 10^4 $ and $\sim 42$,
respectively.  As a result, electrons,
protons, and even ionized helium atoms can be accelerated easily along the
magnetic field lines by parallel electric field presented in the
polar caps of neutron stars in LMXBs.   Since
the magnetic field lines in the
polar cap regions are not closed and extended
beyond the star's light cylinder, the accelerated particles in these
regions may leave the neutron star magnetosphere.  The resulting
flow of particles from the polar caps to infinity produces
a relativistic ``stellar wind'', that exerts a ``stellar wind torque''
\citep{Mic91} on the shell, and then causes
the net angular momentum of the shell to change during the burst.
The net wind torque exerted on the shell due to the out flowing charged
particles from the star polar caps is estimated by \cite{Mic91} for
the Goldreich-Julian charge density to be
\st\be\label{T}
T_W= I^2/4c\Omega_s
\ee
where $I=\eta n_{\rm GJ}e v A_{\rm pc}$ is the total current of the 
charged particles flowing from the polar caps with the area 
$A_{\rm pc}\approx 1.6\pi R^3/R_c$ that is $10$\%
of total star's area ($4\pi R^2$).
Here $n_{\rm GJ}\simeq 0.36\Omega_s B/2\pi e c \simeq
2.25\times 10^{18} (\nu_s/300~{\rm Hz})(B/10^8~{\rm G})$ cm$^{-3}$
is the Goldreich-Julian charge number density, $e=1.6\times 10^{-19}$ C, 
$R_c=(GM_s/\Omega_s^2)^{1/3}$ is the corotation distance
for a star with mass $M_s$, and $v\geq 0.1 c$ is average speed of
particles at $z\sim 10^{-3}$. The velocity of particle grows up quickly
to $\sim c$ at $z\sim 0.1$ as it is accelerated by $E_{||}$.
$\eta \geq 1$ is a dimensionless factor (free parameter of the model) such that
the quantity $\eta n_{\rm GJ}$ represents the actual number density
of particles that accelerated and left the star due to
the parallel electric field.
The value of $\eta\simeq 1$ for old and isolated neutron stars (with no
observed bursts), i.e. the
space charge density is nearly Goldreich-Julian charge density for these
stars.  But in neutron stars in LMXBs with an ocean of accumulated
hot matters,
the value of $\eta$ must be greater than $1$, due to the free ejection of 
particles from the front surface of the ocean.  Further, 
during bursting the accumulated materials with $T\geq 10^8$ K on the 
neutron star surface ,
the kinetic energy of particles inside the ocean abruptly increases and 
then,
the space charge density would increase significantly in regions
with non-zero parallel electric field.  As a result, one would
expect that the number particles that may leave the shell would
increase during the burst.

It is necessary to note that
for neutron stars in LMXBs with accretion rate, 
$10^{-11} M_\odot$ yr$^{-1}< \dot{M} < 10^{-8} M_\odot$ yr$^{-1}$,
the existence of pulsar wind torque might not be consistent with accretion.  
One would expect the motion of the accreting material 
to oppose and shut off the radio pulsar wind, and contribute to
the angular momentum of the
system.   This scenario is unlikely ``during the X-ray burst"
due to the following reasons:
huge thermonuclear explosion on the surface of the star causes 
expolsive eruption the accumulated materials from the surface during
the burst with a mass bulk velocity
close to the speed of light.
As a result, one would expect the action of accretion to
more or less shut off in this period, e.g
by evacuation of part of the inner accretion disk during radius expansion.
Furthermore, 
it has been suggested the action of
accretion torque is balanced by the gravitational radiation torque 
in neutron stars in LMXBs, see \cite{B98a,B02} 
for details. Therefore, the increased angular momentum due to accretion is lost to gravitational 
radiation.   The latter mechanism provides a natural explanation for the rotation  
of stars in a narrow range of frequencies, $\nu_s\sim 300$ Hz.
Finally, the magnitude of accretion torque 
$T_a\sim {\dot M}(G M_s R)^{1/2}\simeq 8.6 \times 10^{32}~ {\rm dyne~ cm}~({\dot M}/10^{-9} M_\odot {\rm yr}^{-1})$ 
(for $M_s=1.4 M_\odot$ and $R=10$ km) is one order of magnitude smaller than that of 
the wind torque $T_w\sim 9.32\times 10^{33}$ dyn cm for $\eta \sim 1000$ and $B\sim10^8$ G, 
see Eq. (\ref{T1}) and the discussion below Eq. (\ref{Om-r}).

By replacing $I$ in Eq. (\ref{T}) we obtain
\st\bea\label{T1}
T_w\simeq 9.32\times 10^{27}~{\rm dyne~cm}~\eta^2
\lp\frac{M_s}{1.4M_\odot}\rp^{-2/3}\times\no\\
\hspace{1cm}\lp\frac{\nu_s}{300~ {\rm Hz}}\rp^{7/3}
\lp\frac{R}{10~{\rm km}}\rp^6\lp\frac{B}{10^8~{\rm G}}\rp^2.
\eea
Equation (\ref{T1}) gives the exerted torque on the neutron star surface
due to the flowing charged particles from the pulsar polar caps.
Therefore, the burning shell of the neutron star will   
experience this torque as it expands during the burst.
The angular momentum of the shell at a distance $R$ from the center
of the star in Newtonian dynamics can be written as
\st\be\label{Ang}
\ell=\kappa M R^2 \Omega,
\ee
where constant $\kappa=(I/MR^2)_{\rm shell}$ depends on
the equation of state and $\kappa\leq 2/3$ (an uniform spherical shell).
Here $M$ and $I$ are the mass and moment inertia of the shell.
Equation (\ref{Ang}) may be corrected by
$\ell=\kappa M R^2 \sin^2\theta \Omega_s/ (1-2GM_s/R)^{1/2}$ for a 
relativistic spherical star and by 
$\ell=\kappa M R^2 \sin^2\theta (\Omega_s-\omega)/ (1-2GM_s/R)^{1/2}$
for a rotating relativistic star \citep[see][]{Cum01}.
Here $\omega=J_s/R^3$ where $J_s$ is the total angular momentum of the
star.    However, as shown by \cite{Cum01}, the relativistic corrections
have small contributions to the final results, and so we ignore them.
The time evolution of the angular momentum of the shell will be 
\st\be\label{Ang-Evo}
\frac{\Delta}{\Delta t}( R^2 \Omega)_{\rm shell}
= - T_w/\kappa M,
\ee
or 
\st\be\label{Ang1}
(R^2 \Omega)_{\rm shell} ( 2\Delta R/R +
\Delta\Omega/\Omega)_{\rm shell} = - (T_w/\kappa M) \Delta t. 
\ee
Therefore, the change in the angular
velocity of the burning shell will be
\st\be\label{Om}
(\Delta\Omega/\Omega)_{\rm shell} =  - 2\Delta R /R  -
       (T_w/\kappa M R^2 \Omega)~ \Delta t.
\ee
Here we assumed that the change in the mass of the shell is order of or
less than $(\Delta R/R)^2$ and then we neglect it \citep[see][]{Cum01}.
Equation (\ref{Om}) represent the change in angular velocity of the shell
during the burst.  Now we discuss the change in angular
velocity in three stages,
before the X-ray burst or the recurrence period, the rising period,
and after the burst.

\subsection{Before the X-ray burst}\label{bb}

In this period the star accumulates materials in the ocean that
more or less rigidly corotates with the star \citep{Cum01}.
The parallel electric field inside the ocean
is nearly neutralized due to redistribution of charge particles of the 
ocean.  
So in this period, the electric field more or less start from zero above
the front surface of the shell and increase by height due to the
charge deficit caused by free emission of charged particles from the
front surface.   The space charge
density above the front surface would be bigger than but not to far
from the
Goldreich-Julian charge density and increases by time due to the
thermonuclear activities in the oceans. So we expect that
$\eta\sim 10-100$.

However, the polar cap acceleration is more or less suppressed due
to strongly accreting plasma in this period.   As a result, the pulsar
wind torque would be negligible.

Furthermore, because of nearly rigid rotation of the ocean with star the net wind
torque
during this period will be acted 
more or less on the star itself.  Therefore, Eq. (\ref{Om}) should be
corrected for this period as 
\st\bea\label{Om-b}
\lp\frac{\Delta\Omega}{\Omega}\rp_{\rm shell} &=& 
 - (T_w/\kappa' M_s R^2 \Omega) ~\Delta t,\no\\
&\simeq&  - 4.44\times 10^{-17} 
\lp\frac{\eta}{100}\rp^2
\lp\frac{M_s}{1.4 M_\odot}\rp^{-5/3}\no\\
&&\hspace{-1.6cm}\times\lp\frac{R}{10~{\rm km}}\rp^{4}
\lp\frac{\nu_s}{300~ {\rm Hz}}\rp^{4/3}
\lp\frac{B}{10^7~ {\rm G}}\rp^{2}
\lp\frac{\Delta t_{\rm rec}}{10^4~{\rm s}}\rp\,,
\eea
where $\Delta t_{\rm rec}$ is the burst recurrence time and $\kappa'=2/5$.
As is clear, the net wind torque on the shell/star in this period and so
the change in angular velocity of the shell/star is negligible.

\subsection{The rising time}

This period starts with abruptly thermonuclear ignition of the accumulated 
hydrogens and heliums during the recurrence time.  The burning 
layers expand in less than $\sim 1$ second up to $z\simeq 2\times 10^{-3}$ 
\citep{AJ82,HF84,B98}.   In this period, as in the hydrostatic models,
we assume that the rotational evolution of the expanding shell is nearly
independent of the star itself. 

As we mentioned before, in this period due to the huge thermonuclear
explosion on the surface of the star,
the accumulated materials are explosively erupted out of the star 
with a mass bulk velocity close to the speed of light.
The outgoing plasma density within period of time $\Delta t\sim 1$ second, is 
$\sim [(A_{\rm pc}/4\pi R^2) M_{\rm shell}]/A_{\rm pc}$ that is $10$ times
(or more) larger than in-falling plasma density $m_{\rm acc}/A_{\rm pc}$
due to the accretion flow at the same period of time.   Here
$m_{\rm acc}={\dot M}\Delta t$ is the average mass of accreting
materials during $\Delta t$.
As a result, one would expect the action of accretion to
more or less shut off during this period.

The parallel electric force at height $z\simeq 2\times 10^{-3}$
is $80$ times bigger than gravitational force for hydrogen atoms,
see Eq. (\ref{E||}).  Therefore, the charged particles that rise with 
the expanding layers will be accelerated toward the outside of light 
cylinder by the strong parallel electric field along the open magnetic field 
lines.
Further, the sudden thermonuclear explosion of the accumulated materials
in the ocean causes the thermal energy of the charged particles to increase.  
Consequently, we expect that the number density of particles that
may leave the magnetosphere $\eta n_{\rm GJ}$ in this period 
increases significantly.
The resulting wind torque for $\eta\sim 10^3$ will give a
net change in angular velocity of the shell as
\st\bea\label{Om-r}
\lp\frac{\Delta\Omega}{\Omega}\rp_{\rm shell} &=& 
-2|\Delta R|/R - (T_w/\kappa M R^2 \Omega) ~\Delta t,\no\\
&\simeq& -2\times 10^{-3} \lp\frac{z}{10^{-3}}\rp\no\\
&&\hspace{-1.2cm} -~ 7.42\times 10^{-3}
\lp\frac{\eta}{10^3}\rp^2
\lp\frac{M_s}{1.4 M_\odot}\rp^{-2/3}
\lp\frac{M}{10^{21}~{\rm g}}\rp^{-1}\no\\
&&\hspace{-1.4cm}\times\lp\frac{R}{10~{\rm km}}\rp^{4}
\lp\frac{\nu_s}{300~ {\rm Hz}}\rp^{4/3}
\lp\frac{B}{10^8~ {\rm G}}\rp^{2}
\lp\frac{\Delta t_{\rm rise}}{1~{\rm s}}\rp,
\eea
where $\kappa=2/3$ and $\Delta R\simeq 10$ m is the average change
in thickness of the burning layers during a burst. 
For $\eta\sim 10^3$ and $B\sim 10^8$ G, the net change in angular
velocity of the shell given by Eq. (\ref{Om-r})
is in good agreement with large frequency shifts seen in observations.
We note that for $\eta\sim 10^3$, the number of particles that left the
star $\Delta N\sim 4\pi\eta n_{\rm GJ}R^2\Delta R$ is
much smaller than total number of particles in the ocean 
$N\sim M/m_p$, i.e. $\Delta N/N\simeq 4\times 10^{-7}$.   As expected, 
the change in mass of the shell is order of $(\Delta R/R)^2\sim 10^{-6}$.
In Table 1 we calculate the quantity $\eta B$ for both $300$ Hz and
$600$ Hz spin frequencies (for those which
applicable), to produce the frequency shifts observed in various systems.
As a result, the suggested magnetic fields are in range
$1-10 \times 10^7$ G for $\eta\sim 100-1000$.  This is consistent
with other observational evidences about the magnetic field of 
neutron stars in LMXBs.

\subsection{After the X-ray burst}

At final stage, after the thermonuclear flash,
the expansion will be stopped by the star's gravity force,
and then the expanded layers start to contract as they cool down.
Because of the charge redistributions
during the expansion, the original induced electric field
is neutralized inside the expanded shell.
So the magnitude of the parallel electric field inside the ocean
will drop to zero, and then the electrostatic acceleration
will cease inside the ocean.    Due to the lack of outward
electrostatic
acceleration, the ocean's particles are mostly accelerated downward by
strong gravity.
Furthermore, the action of accretion plasma is starting to recover,
and as a result, the polar cap acceleration will shut down
in this period.
Although, because of new charge distribution, a new parallel electric
field may develop above the burned front
we expect that
the polar cap particle acceleration mechanism is more or less unlikely
after the burst.   As a result, the number of particles that may leave the
star will decrease as well, i.e. $\eta\leq 10$.  
Further as the shell cools down, it gets closer and closer to the
star, and then its coupling to the star gets stronger and stronger.
This can be understood by noting the magnetic field near the star's
surface opposes the differential rotation arising between shell's layers
and surface of the star during the burst.
Therefore, the net wind torque will be acting more and more
on the whole star again rather than shell itself, and so we neglect it,
see Sect. \ref{bb}.

The magnetic recoupling of the shell forces the shell to
spin up until it achieves the star's spin frequency.
The shell gains its angular momentum deficit
$(MR^2\Delta\Omega)_{\rm shell}$, from the star,
and then it reduces the angular velocity of the star by
$\delta\Omega_s = (M/M_s) \Delta\Omega
\simeq 10^{-12} \Delta\Omega$ which is negligible.
As a result, one would expect that the oscillations' frequency
will increase as the shell spins up by
$\Delta\nu/\nu_0\sim \Delta\Omega/\Omega$.
In Fig. 1, we compare the corresponding changes suggested by previous
studies \citep{Str97,Cum01} and the one we discussed here.  
As is clear, due to the exerted torque by the particle wind in the polar cap 
regions, the shell's spin frequency,
and then the oscillations' frequency decrease more than in the
calculation by \cite{Cum01} during the rising time.   
Therefore, we expect to observe larger change in
oscillations' frequency during the burst tail.


\section{Discussion}\label{discuss}

Among the $\sim 50$
known Galactic LMXBs, the highly coherent burst oscillations with
large modulation amplitudes and stable frequencies in range
$\nu_0\sim 270 - 620$ Hz are seen in ten LMXBs
during Type I X-ray bursts.  These oscillations
are most commonly seen during tails of bursts,
when the burning is thought to have spread over the whole surface and
obviously asymmetry is no longer present, are not observed
in all Type I X-ray bursts from the same source.
While it is believed that the oscillation frequency is closely related
to the neutron star spin frequency due to the observation of 
the kHz quasi-periodic oscillations in the persistent
emission \citep{Kli00}, there is still an ambiguity that 
whether the oscillation frequency is the spin
frequency or twice the spin frequency.
Further, as seen in observations, the oscillation
frequency increases by $\Delta \nu$ a few Hz during the burst.
This frequency shift is firstly explained by \cite{Str97} that
the burning shell decouples from the star, and undergoes spin changes 
due to the conservation of angular momentum of the shell
as it expands and contracts during
the Type I X-ray bursts.  Further observations and studies 
suggested that purely radial hydrostatic expansion
and angular momentum conservation alone cannot explain
rather large frequency drifts ($\Delta\nu/\nu\sim 1.3$\%) observed  
in some bursts \citep{Cum01,Gal01,Wij01}.   For recent review on the
thermonuclear bursts and their properties see \cite{SB03}.

In this paper, we addressed the latter problem by studying the evolution
of angular momentum of the burning
shell during the Type I X-ray bursts in LMXBs. 
Based on particle acceleration models near a pulsar
polar cap region \citep{Mes98},
we studied the change in the angular momentum of the burning shell
during the Type I X-ray burst.  
The net charged particles that accelerated by parallel electric
field in star's polar caps, flow from the star
surface to infinity through open magnetic field lines, 
would exert a torque on the burning shell (called wind torque) 
and cause the angular momentum of the shell changes during the burst.
We introduced a dimensionless factor $\eta$ that the quantity
$\eta n_{\rm GJ}$ represent the number density of the ejected particles
from the burning layers above the pulsar polar caps. 
We showed that for $\eta\sim 100-1000$ (during X-ray bursts)
and a typical magnetic field $B\sim 1-10\times 10^7$ G in 
the pulsar polar caps, the rather large observed
frequency drifts of burst oscillations can be explained by 
the resulting wind torque exerted on the burning shell. 
In Table 1, we obtained the values for $\eta B$ for both $300$ Hz and
$600$ Hz spin frequencies 
that causes the corresponding frequency drift observed in each burst.
As is clear, the resulting magnetic fields' strength is in a good agreement
with other observations from the neutron stars in LMXBs, 
such as the lack of coherence pulsations in persistent emission. 

Above the neutron star's polar caps particles flow outward along the
open magnetic field lines and a steady charge density cannot be
maintained at the Goldreich-Julian density everywhere.   Consequently,
a strong electric field develops along the magnetic field, extracts
the `primary' charged particles with density $n\sim n_{\rm GJ}$ (for a 
star with no burst activity, see Sect. \ref{model}) 
from the surface and accelerate them to high Lorentz factor
($\sim 10^7$).   
According to the polar cap models, the predicted radiation luminosity for
a typical millisecond pulsar would be
${\cal E}\sim 10^{33} (\nu_s/300~{\rm Hz})^{5/28}(B/10^8~{\rm G})^{-1/7}$
ergs/s that is comparable to $\gamma$-ray luminosities.
So, one may expect to observe $\gamma$-rays in millisecond pulsars that
has been detected only in PSR J0218+4232 by EGRET \citep{Kui00}.
However, polar cap particle acceleration is limited by  
various energy loss mechanisms such as curvature radiation,
resonant and nonresonant
inverse Compton scattering, and pair production.  The accelerated
particles along the magnetic field lines suffer energy loss and
emit photons through the curvature radiation. Further, these particles
scattered by soft thermal X-ray photons emitted from the hot surface and
loss energy via inverse Compton scattering.  In addition, the high energy
photons resulted from the curvature radiation and/or inverse Compton
scattering produce a cascade of `secondary' $e^\pm$ pair particles
with a number density $n_{e^\pm} \sim 10^5 n$
above polar cap region that change the charge distributions and screen
the electric field \citep[see for more detail][]{HM01}.
In millisecond pulsars (with no burst activities) the
polar cap particle acceleration mechanism was studied by \cite{LSM00}.
They showed that in these pulsars this mechanism, which is relatively
efficient in accelerating charged particles in the star's polar caps,   
mostly suffers energy 
loss from the radiation of accelerated particles along the field lines (
curvature radiation).  So  
the maximum energy attainable by the particle is
limited by energy loss through the radiation reaction.
They found that the maximum Lorentz factor that might be achieved
by an individual particle at $z\sim 0.1$  
is $\Gamma= (6\pi\varepsilon_0 E_{||}\rho_c^2/e)^{1/4}\sim 10^7$, however,
the Lorentz factor for the bulk pair plasma is much smaller $\sim 100$.  
The latter may explain the lack of detection of high energy
$\gamma$-rays in these stars. Here $E_{||}$ is given by Eq. (\ref{E||}) and
$\rho_c\sim (4/3)(cR/2\pi\nu_s)^{1/2}\sim 5\times 10^4
(300~{\rm Hz}/\nu_s)^{1/2} (R/10^6~{\rm cm})$ m is
the curvature radius of the field lines.

For neutron stars in LMXBs with intermediate accretion rate $10^{-11} M_\odot$ yr$^{-1}< \dot{M} < 10^{-8} M_\odot$ yr$^{-1}$, however,
the polar cap acceleration mechanism mostly suffers from the action of
accretion plasma.   Because of the large plasma density ($m_{\rm acc}/A_{\rm pc}$)  
of the accretion flow toward star's polar caps, the parallel electric field in the 
polar cap regions would be shorted out 
and the polar cap acceleration mechanism would be suppressed.   We note that, this 
scenario is not likely during X-ray bursts due to the explosively 
erupting materials from the surface and possible evacuation of part of 
inner disk.  In this period with $\Delta t\sim 1$ second,
the eruptive plasma density is $10$ times larger that the accreting plasma 
density, ie. 
$\sim [(A_{\rm pc}/4\pi R^2) M_{\rm shell}]/A_{\rm pc}\sim 10~
m_{\rm acc}/A_{\rm pc}$.

\acknowledgements
The author is grateful to Sharon M. Morsink and Supratim Sengupta
for reading the manuscript and stimulating discussions during the
course of this work.  
I would like also thank to the referee for his/her valuable comments.  
This research was supported by the
Natural Sciences and Engineering Research Council of Canada.

\vspace{2cm}
%
\noindent {\bf Table Captions:}\\
\begin{itemize}
\item{Table 1:} 
In this table we obtain the lowest radius expansion $\Delta R$ (based on
hydrostatic expansion models) and the scaled magnetic field $\eta B$
(based on the polar cap
particle acceleration models) for some X-ray bursts in LMXBs.
We calculate these quantities for two chosen values of the star
spin frequency, $\nu_s=300$ Hz and $\nu_s=600$ Hz, 
such that the corresponding frequency shift is comparable with observations.   
$\nu_0$ is the oscillation frequency and $\Delta \nu$ its corresponding
shift seen during the burst.  The value of $\eta B$ is based on the
assumed value of $\Delta R\simeq 20$ m for both spin frequencies,
see Eq. (\ref{Om-r}).
\end{itemize}
\begin{center}
Table 1\vspace{.52cm}\\
\begin{tabular}{cccccccccc}\hline\hline
Object & Time &  $\nu_0$ & $\Delta\nu/\nu_0$ &
$\Delta R_{300}$ & $\Delta R_{600}$ & $(\eta B)_{300}$ & $(\eta B)_{600}$\\
       & (UT) &  (Hz) &($10^{-3}$) &
       (m) & (m) &  ($10^{10}$ G) & ($10^{10}$ G)
        \\\hline
4U 1636-54 & 1996 Dec 28 (22:39:22)$^{1,2}$ & 580.5 & $\sim$ 2-4
& $\geq$ 40 & $\geq$ 20 & $\geq$ 3 & $\geq$ 1 \\
           & 1996 Dec 29 (23:26:46)$^{2,3}$ &  581.5 & $\sim$ 3
& $\geq$ 40 & $\geq$ 20& $\sim$ 2 & $\leq$ 0.01 \\
           & 1996 Dec 31 (17:36:52)$^{2,3}$ &  581. & $\sim$ 3
& $\geq$ 40 & $\geq$ 20& $\sim$ 2 & $\leq$ 0.01 \\
4U 1702-43 & 1997 Jul 26 (14:04:19)$^{4~~~}$ &  329.85 $\pm$ 0.1
& 7.7 $\pm$ 0.3 & $\geq$ 50 & - & $\geq$ 6 & -\\
           & 1997 Jul 30 (12:11:58)$^{4~~~}$ &  330.55 $\pm$ 0.02
& 4.8 $\pm$ 0.3 & $\geq$ 30 & - & $\geq$ 2 & - \\
4U 1728-34 & 1996 Feb 16 (10:00:45)$^{4,5}$ &  364.23 $\pm$ 0.05
& 6.6 $\pm$ 0.1 & $\geq$ 50 & - & $\geq$ 5 & - \\
           & 1997 Sep 9 (06:42:56)$^{4~~~~}$ &  364.10 $\pm$ 0.05
& 5.9 $\pm$ 0.2 & $\geq$ 45 & - & $\geq$ 4 & - \\
Aql X-1        & 1997 Mar 1 (23:27:39)$^{6,7~}$ &  549.76 $\pm$ 0.04
& 4.3 & $\geq$ 50 & $\geq$ 25 & $\geq$ 3 & $\sim$ 1 \\
4U 1916-053 & 1998 Aug 1 (18:23:45)$^{8~~~}$ &  269-272
& 13.0 & $\geq$ 70 & - & $\sim$ 12.5 & -\\
MXB 1658-298 & 1999 Apr 14 (11:44:52)$^{9~~}$ &  567
& 9.0 & $\geq$ 100 & $\geq$ 50 & $\sim$ 10 & $\sim$ 7.5 \\ \hline
\end{tabular}
\end{center}
References: (1) \cite{Str98}; (2) \cite{Mil00}; (3) \cite{Str99};
(4) \cite{SM99}; (5) \cite{Str96};
(6) \cite{Zha98}; (7) \cite{Fox00};
(8) \cite{Gal01}; (9) \cite{Wij01}

\newpage
\begin{figure*}
\centering
\includegraphics[width=15cm]{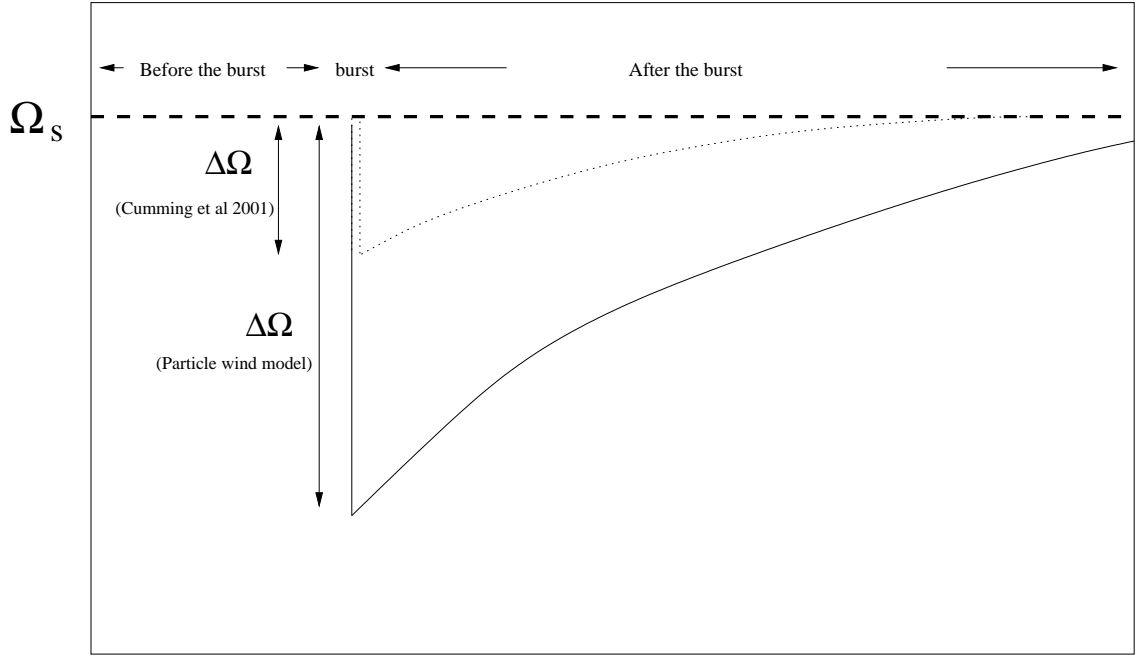}
\caption{
In this figure we schematically compare the change in angular velocity of
the expanding shell and then oscillations' frequency during the type I X-ray
burst suggsted by \cite{Cum01} (dotted line) and the one proposed here
(solid line).   Due to the exerted torque on the shell by the particle wind
from the magnetic polar cap regions during the burst, we expect a
larger depth in $\Delta\Omega$ rather than the model discussed by
\cite{Cum01}. As a result, the larger drift would be expected in the
burst tail.  The solid and dotted curve after the burst are drawn
schematically to show star-shell recoupling in this period.
}
\end{figure*}

\end{document}